\documentclass[aps,prd,preprintnumbers,nofootinbib,twocolumn]{revtex4}
\usepackage{bm}
\usepackage{latexsym}
\usepackage{dcolumn}
\usepackage{amsmath,amsfonts,amssymb}
\usepackage{graphicx,epsfig}
\usepackage{psfrag}
\usepackage{amsthm}

\def\be {\begin{equation}}
\def\ee {\end{equation}}
\def\bea {\begin{eqnarray}}
\def\eea {\end{eqnarray}}
\def\bc {\begin{center}}
\def\ec {\end{center}}
\def\bfg {\begin{figure}}
\def\efg {\end{figure}}
\def\bi {\begin{itemize}}
\def\ei {\end{itemize}}

\def\la {\label}
\def\le {\left}
\def\ri {\right}

\def\no {\noindent}

\def\vs {\vspace}

\def\a  {\alpha}

\def\beq{\begin{equation}}
\def\eeq{\end{equation}}
\def\br{\begin{eqnarray}}
\def\er{\end{eqnarray}}
\newcommand{\eel}[1] {\label{#1}\end{equation}}

\newcommand{\bdm}{\begin{displaymath}}
\newcommand{\edm}{\end{displaymath}}

\begin{document}

\title{Dark matter and dark energy from Bose-Einstein condensate
}

\author{Saurya Das $^1$} \email[email: ]{saurya.das@uleth.ca}
\author{Rajat K. Bhaduri$^{2}$}\email[email: ]{bhaduri@physics.mcmaster.ca}

\vs{0.3cm}

\affiliation{$^1$ Department of Physics and Astronomy,
University of Lethbridge, 4401 University Drive,
Lethbridge, Alberta, Canada T1K 3M4 \\}

\affiliation{$^2$ Department of Physics and Astronomy, McMaster University, Hamilton, Ontario,
Canada L8S 4M1 \\}

\begin{abstract}
We show that Dark Matter consisting of bosons of mass of about $1~eV$ or less
has critical temperature
exceeding the temperature of the universe at all times, and hence would have formed a
Bose-Einstein condensate at very early epochs. We also show that the wavefunction of this condensate,
via the quantum potential it produces, gives rise to a cosmological constant
which may account for the correct dark energy content of our universe.
We argue that massive gravitons or axions are viable candidates for these constituents.
In the far future this condensate is all that remains of our universe.

\end{abstract}

\maketitle


%
%
The basic contents of our universe in terms of Dark Matter (DM),
Dark Energy (DE), visible matter and radiation,
and also its accelerated expansion in recent epochs has been firmly established by a number of
observations now \cite{perlmutter,riess,wmap,bao}.
However although DM constitutes about $25\%$ and DE about $70\%$ of all
matter-energy content, the constituents of DM and the origin
of a tiny cosmological constant, or DE, of the order of $10^{-123}$ in Planck units,
which drives this acceleration, remains to be understood.
In this paper we show that if DM is assumed to consist of a gas of bosons of mass $m$,
then for $m \leq 1 eV$, the critical temperature below which they will
form a Bose-Einstein condensate (BEC) exceeds the temperature of the universe at
all times. Therefore they would form such a condensate at very early epochs, in which
a macroscopic fraction of the bosons fall to the ground state
with little or no momentum and zero pressure,
and therefore may be considered as viable candidates for cold DM (CDM).
Further, via the quantum potential that it produces,
the macroscopic wavefunction of the condensate gives rise to a positive cosmological constant
in the Friedmann equation, whose magnitude depends on $m$, and
for $m \simeq 10^{-32}~eV$, one obtains the observed value of the cosmological constant.
Therefore bosons with this tiny mass can account for both DM and DE in our universe.
We argue that massive gravitons or axions are viable candidates for these bosons.
Finally we speculate on the ultimate fate of our universe, and
end with some open problems.

To compute the critical temperature of an ideal gas of bosons constituting DM, we first note that
these must have a mass, however small, and with
average inter-particle distances $(N/V)^{-1/3}$ (where $N=$ total number of bosons in volume $V$)
comparable or smaller than the thermal de Broglie wavelength $hc/k_B T$, such that
quantum effects start to dominate.
Identifying this temperature of a bosonic gas to the critical temperature $T_c$
(below which the condensate forms) we get $k_B T_c \simeq h c (N/V)^{1/3}$.
A more careful calculation for ultra-relativistic
noninteracting bosons with a tiny mass gives
\cite{brack,grether,fujita}
\footnote{One can also consider a shallow three dimensional harmonic oscillator trapping
potential with angular frequency $\omega$, for which the
Gaussian wavefunction is a coherent ground state,
one has $T_c = (N/\eta)^{1/3} \hbar \omega/k_B$ \cite{brack},
which on using $L_0= \sqrt{\hbar/m\omega}$, $m=h/cL_0$ and the fact that $N/V$ is constant, gives
$T_c = (\hbar c/k_B) \le( N/\eta V \ri)^{1/3}$, virtually identical to Eq.(\ref{tc1})
up to a factor of order unity, lending it further credence.}
\footnote{For previous studies of superfluids and BEC in cosmology see
\cite{sudarshan,morikawa,moffat,wang,boehmer,sikivie,dvali,chavanis,kain,suarez,ebadi,laszlo1,bettoni,gielen,schive,davidson}.
See note at the end for more information.
}
\bea
T_c = \frac{\hbar c}{k_B} \le( \frac{N \pi^2}{V \eta \zeta(3)} \ri)^{1/3}~.
\la{tc1}
\eea
In the above
$N=N_B + N_R$, $N_B$ being the number of bosons in the BEC, and $N_R$ outside it,
both consisting of bosons of small mass as discussed earlier,
$\eta$ is the polarization factor and
$\zeta(3) \approx 1.2$.
Also $a$ is the cosmological scale factor, $L=L_0 a$ is the Hubble radius,
$V=L^3=L_0^3 a^3 =V_0 a^3$
(subscript $0$ here and in subsequent expressions denote current epoch,
when we also assume $a=1$). 
Note that for boson temperature $T<T_c$, a BEC will necessarily form, even when there are interactions
\cite{bhaduricjp}.
As stated before, identifying the BEC of bosons all in their ground states with zero momenta and only
rest energies 
with DM in any epoch, we estimate
$N_B/V= \rho_{DM}/m =0.25 \rho_{crit}/ma^3$
%
i.e. $N \simeq N_B$, and obtain from Eq.(\ref{tc1})
\bea
T_c = \frac{6 \times 10^{-12}}{m^{1/3}~ a}~K~,
\la{tc2}
\eea
($m$ in kg in the above). Therefore if $m<1~eV$, $T_c > 2.7/a$, the universe
background temperature at all times, and
a BEC of the constituent bosons will form at very early epochs.
%
%
As mentioned earlier, when the bosons are in the BEC state, they have little or no
momentum, and behave as CDM.

Furthermore, as is well-known, a BEC is a macroscopic quantum state, described by
a wavefunction $\phi$, which we decompose in the form
$\phi={\cal R}e^{iS}$
(${\cal R} (x^\a),S (x^a)=$ real functions, and $|\phi|^2={\cal R}^2$ represents the
spatial density of the bosons in the BEC).
Next we replace classical geodesics by
quantal (Bohmian) trajectories defined by the velocity field
$u_a=\hbar ~\partial_a S/m$ (as one should, in a quantum mechanical description)
\cite{bohm,holland}, and also define the induced metric
$h_{ab} = g_{ab} - u_a u_b$, using which
it is found that the quantum corrected second order Friedmann equation (Raychaudhuri equation)
for the scale factor $a(t)$ is \cite{sd,dasessay,alidas}
\footnote{see also \cite{alberto1} and \cite{he} for application of Bohmian mechanics in cosmology.}
\bea
\frac{\ddot a}{a} = - \frac{4\pi G}{3} \le( \rho + 3p \ri)
 + \frac{\hbar^2}{3 m^2} h^{ab} \le( \frac{\Box {\cal R}}{\cal R} \ri)_{;a;b}
~,
\la{frw1}
\eea
where $\rho=\rho_{vis}+\rho_{DM}$ is the sum of densities from visible matter and DM
(similarly for the pressure).
%
We interpret the last ${\cal O}(\hbar^2)$ term as the quantum mechanical
contribution to the cosmological constant
\bea
\Lambda_Q = \frac{\hbar^2}{m^2 c^2} h^{ab} \le( \frac{\Box {\cal R}}{\cal R} \ri)_{;a;b}~.
\la{qlambda}
\eea
In the above, $V_Q \equiv (\hbar^2/m^2 c^2) \Box {\cal R}/{\cal R}$ is known as the
relativistic {\it quantum potential} \cite{sd}.
Note that $V_Q$ or $\Lambda_Q$ are not ad-hoc, but in fact always present
in a quantum description of the contents of our universe,
and vanishes in the $\hbar \rightarrow 0$ limit.
From Eq.(\ref{qlambda}) it follows that $\Lambda_Q$ depends on the amplitude ${\cal R}$ of the
wavefunction $\phi$, which we take to be the ground state of a
condensate. Note that with this choice, ${\cal R}$ is also
time-independent.
Its exact form is not important to our argument however, except that it is non-zero
and spread out uniformly over the range $L_0$ of the observable universe, or the Hubble radius,
with minute non-uniformities present at much smaller scales.
This follows from the cosmological principle (homogeneity and isotropy of our universe)
as well as from causality, the latter requiring that anything outside the Hubble radius would not influence the accessible
wavefunction. Also as shown in \cite{wald} modes with wavelengths greater than this radius
decay rapidly.
Thus either using a straightforward dimensional argument, or a generic wavefunction such as a
Gaussian with a large spread ${\cal R} ={\cal R}_0 \exp(-r^2/L_0^2)$,
which is also the ground state for a shallow three dimensional
harmonic oscillator potential \cite{brack},
or one which results when an interaction of strength $g$ is included such that
${\cal R} = {\cal R}_0 \tanh(r/L_0\sqrt{2})~(g>0)$ and
${\cal R} = \sqrt{2}~{\cal R}_0~\mbox{sech} (r/L_0)~(g<0)$ \cite{rogel}, one obtains
$\le( \Box {\cal R}/{\cal R}\ri)_{;a;b} \simeq 1/L_0^4.$
Furthermore from quantum mechanics, $L_0$ also determines the characteristic range of the wavefunction $\phi$,
and as such may be identified with the Compton wavelength of the bosons of mass $m$ under consideration,
i.e. $L_0=h/mc$ \cite{wachter}, from which it follows
%
\bea
\Lambda_Q = \frac{1}{L_0^2} =\le( \frac{m c}{h} \ri)^2 ~, \la{lambda1}
\eea
Thus for the current Hubble radius $L_0 = 1.4 \times 10^{26}~metre$,
one obtains $m \simeq 10^{-68}~kg$ or $10^{-32}~eV$,
%
%
%
Finally, inserting the above value of $L_0$ or $m$ in Eq.(\ref{lambda1}), one obtains
\bea
\Lambda_Q && = 10^{-52}~(metre)^{-2} \\
&& = 10^{-123}~\ell_{Pl}^{-2}~(\text{Planck units})~,
\eea
where $\ell_{Pl} = 1.6 \times 10^{-35}~metre$ is the Planck length.
This entirely accounts for the observed value of the cosmological constant,
without the need to put it in the Friedmann equation by hand.
We therefore see that bosons of such tiny mass in a BEC can
account for both the DM (via its density) and DE (via quantum potential potential of its macroscopic
wavefunction).
Also in this case, from Eq.(\ref{tc2}), the critical temperature
is $T_c = 10^{11}~a^{-1}~K$
much higher than the universe temperature at all times
as stated earlier, confirming that the BEC will indeed form at the very early
stages of our universe.

But what could be these bosons constituting DM and giving rise to DE?
There are at least two viable candidates.
First, we consider massive gravitons. Although gravitons derived from general relativity
are massless, there has been considerable progress recently, both in the theoretical and
experimental fronts, in having a consistent picture of massive gravitons in extensions of
general relativity.
For example, on the theoretical side,
as early as in the 1930s gravitons of  mass ${\cal O}(10^{-32})~eV$ had been proposed \cite{zwicky}.
More recently it was shown that
massive gravitons can appear due to spontaneous symmetry breaking \cite{mukhanov,oda},
in a completely covariant non-linear completion of the Fierz-Pauli type
massive gravity action \cite{derham}, and in ghost free theories \cite{hassan},
solving an age old problem of having a covariant theory of massive gravitons.
These theories also clearly admit graviton mass ${\cal O}(10^{-32})~eV$.
Also as was recently shown, study of cosmology within these theories gives rise to additional
densities in the Friedmann equation, which can be included in our
definition of $\rho$ in Eq.(\ref{frw1}) \cite{massive1,massive2,derham2}.
Other theoretical approaches also point to graviton mass in this range,
\cite{mann,hinterbichler,majid}.
In the Newtonian limit,
for gravitons of mass $m$, the corresponding gravitational field follows a Yukawa type of force law
$
F  \propto
\frac{k}{r^2}~e^{-r/L_0}~.
$
Since gravity has not been tested beyond this length scale,
such an interpretation is natural and cannot be ruled out.
Also if one invokes periodic boundary conditions, this is also the mass of the lowest
Kaluza-Klein modes.
All bounds on graviton masses obtained from observations too
suggest graviton mass ${\cal O}(10^{-32})~eV$ \cite{graviton}.
Finally, gravitons as DM during the inflationary stage may also have
observational consequences at present \cite{anupam}.

The second possibility is axions. Although axions were originally proposed
to solve the strong CP problem in
quantum chromodynamics, they also arise in the context of string theory,
and have long been advocated as DM candidates \cite{axion}, and BEC of axionic DM have also been
explored \cite{sikivie}.
Axion mass depends on the form of the action considered and couplings therein, but
masses ${\cal O}(10^{-32})~eV$ are certaintly not ruled out \cite{marsh,marsh2}.
Experiments to detect axionic DM are also in progress \cite{admx}.
It must be kept in mind however that until detected, axions remain as hypothetical
particles, requiring extension of the otherwise well-tested standard model of particle physics,
with their masses and couplings put in by hand.

As for the dynamics of DM, the BEC wavefunction and its fluctuations should be governed
by the Gross-Pitaevskii equation (GPE) or its relativistic generalization
\cite{pethik,liberati}
\bea
\le[ \Box + m^2 + g|\phi|^2   \ri]\phi = 0~,
\eea
which in a Hubble background, and for negligible self-interactions ($g \ll 1$)
is consistent with the dynamical equation (13) of \cite{marsh}
and perturbation equations derived from it.
It may be noted that the above equation holds for a BEC of gravitons as well.
It has also been shown
that at galaxy length scales, a BEC of light particles
naturally gives rise to DM density profiles which match well with observed
galaxy rotation curve velocities (see e.g. \cite{wang,boehmer,laszlo1}).

At late times with universe temperature $T \ll T_c$, one has \cite{brack}
\bea
\frac{N_B}{N} = 1 - \le( \frac{T}{T_c} \ri)^3 \rightarrow 1~,
\la{ratio1}
\eea
i.e. $N_B \rightarrow N$, $N_R \rightarrow 0$, and most available bosons
would be subsumed within the condensate, accounting for the DM in our universe.
%
%
The density of the latter of course falls off as $1/a^3$, while that of
DE (via the $\Lambda_Q$ term) remains constant.
It follows that in the end the latter is all that remains, just as predicted by the DE hypothesis,
with the universe being described by a giant quantum state of the condensate.
This situation is depicted in Figure 1, where one can see that
the BEC dominates at later times and lower background temperatures, when
as we argued, the condensate provides a viable source of DE and DM.
This may also be one of the best evidences for the graviton \cite{krauss}.

In summary, we have shown here that bosons of tiny mass should form a BEC
at early times and may account for the DM content of the universe, while
its macroscopic wavefunction can account for the DE.
Both massive gravitons and axions are viable candidates of bosons of
this mass, and hence that of DM and DE.
While the former requires a modification of general
relativity, the latter requires extension of the standard model.
While this picture predicts a high degree of homogeneity
and isotropy at large scales (as observed),
it still allows for relatively small variations of densities, temperatures etc.
at smaller scales.
It will be interesting to investigate other testable predictions of this BEC, such as its heat capacity,
the distribution of DM, response to galaxy rotations etc.
We hope to report on these elsewhere.

\vspace{0.2cm}
\noindent
{\bf Note added} \\
In this note we list features of previous works in BEC in the context of cosmology.
In \cite{sudarshan}, the authors use the
same formula for $T_c$ as our Eq.(\ref{tc1}), and conjectured that the infinite heat conductivity of the
BEC may account for the uniform microwave background temperature.
In \cite{morikawa}, it was proposed that the BEC manifests itself as DE and DM at different epochs.
In \cite{moffat}, DM candidate of a BEC formed out of the `phion' field in modified gravity was considered.
In \cite{wang}, a cold star composed of a dilute BEC was studied.
In \cite{boehmer}, \cite{chavanis}, \cite{kain}, \cite{suarez}, \cite{laszlo1} and \cite{bettoni},
properties of various forms of BEC DM were studied.
In \cite{sikivie} and \cite{davidson}, the possibility of axions as a BEC was examined.
Background geometries and black holes made up of BEC was considered in \cite{dvali}.
DM composed of a BEC of particles obeying infinite statistics was studied in \cite{ebadi}.
In \cite{gielen} BEC in loop quantum cosmology was studied.
In \cite{schive} it was shown that DM can be well approximated by BEC at large scales, although their estimate of
boson mass was higher than that proposed in this paper. 
%

\vs{0.9cm}
\begin{center}
\begin{figure}
\includegraphics[scale=0.55,angle=0]{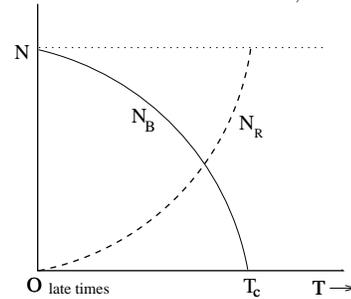}
\caption{$N_B$ and $N_R$ vs. $T$.}.
\label{solution}
\end{figure}
\end{center}

\no {\bf Acknowledgment}

\no
SD thanks L. A. Gergely and M. W. Hossain for discussions.
This work is supported by the Natural Sciences and Engineering
Research Council of Canada.
%



\end{document}